\documentclass[aps,prb,twocolumn,amsmath,amssymb,floatfix,footinbib,bibnotes,groupedaddress,longbibliography]{revtex4-1}

\usepackage{soul}
\usepackage{bm}
\usepackage{epsf}
\usepackage{amssymb,slashed}
\usepackage{esint}
\usepackage{amsmath}
\usepackage{comment}
\usepackage{widetable}
\usepackage{graphicx}
\usepackage{tabularx}
\usepackage{rotating}
\usepackage{epsfig}
\usepackage{psfrag}
\usepackage{amsmath}
\usepackage{subfigure}
\usepackage{nicefrac}
\usepackage{bm}% bold math
\usepackage[usenames,dvipsnames]{color}
\usepackage{soul}
\usepackage{multirow}
\usepackage{titlesec}
\usepackage{mathtools}
\usepackage[colorlinks,citecolor=blue,urlcolor=blue,linkcolor=blue]{hyperref}
\newcommand{\vect}[1]{\boldsymbol{#1}}
\newcommand{\vecta}[1]{\boldsymbol{#1}}
\newcommand{\vectb}[1]{#1_\mu}

\newcommand{\textfrac}[2]
{{\textstyle\frac{#1}{#2}}}
\newcommand*\xbar[1]{%
 \hbox{%
 \vbox{%
 \hrule height 0.3pt % The actual bar
 \kern0.3ex% % Distance between bar and symbol
 \hbox{%
 \kern-0.1em% % Shortening on the left side
 \ensuremath{#1}%
 \kern-0.1em% % Shortening on the right side
 }%
 }%
 }%
}

\def \sgn#1{\text{sgn}\left( #1 \right)}
\def \up{\uparrow}
\def \down{\downarrow}

\global \def \aa#1 {\begin{align} #1 \end{align}}

\newcommand {\apgt} {\ {\raise-.5ex\hbox{$\buildrel>\over\sim$}}\ }
\newcommand {\aplt} {\ {\raise-.5ex\hbox{$\buildrel<\over\sim$}}\ }

\begin{document}

\title{Spin-charge separation in two dimensions: spinon-chargon gauge theories from duality}

\author{Eyal Leviatan}
\affiliation{Department of Condensed Matter Physics, Weizmann Institute of Science, Rehovot 7610001, Israel}

\author{David F. Mross}
\affiliation{Department of Condensed Matter Physics, Weizmann Institute of Science, Rehovot 7610001, Israel}

\begin{abstract}
Strong interactions between electrons in two dimensions can realize phases where their spins and charges separate. We capture this phenomenon within a dual formulation. Focusing on square lattices, we analyze the long-wavelength structure of vortices when the microscopic particles -- electrons or spinful bosons -- are near half-filling. These conditions lead to a compact gauge theory of spinons and chargons, which arise as the fundamental topological defects of the low-energy vortices. The gauge theory formulation is particularly suitable for studying numerous exotic phases and transitions. We support the general analysis by an exact implementation of the duality on a coupled-wire array. Finally, we demonstrate how the latter can be exploited to construct parent Hamiltonians for fractional phases and their transitions.
\end{abstract}

\maketitle

\section{Introduction}
\label{sec.Intro}
Electrons have charge $e$ and spin $\tfrac{1}{2}$ -- this fundamental property determines the possible excitations of most condensed-matter systems. The basic quasiparticles of most metals and insulators carry the same quantum numbers and may thus be viewed as `dressed' electrons. In magnets, the charge degrees of freedom are frozen at low energies, and the elementary quasiparticles are quantized spin waves. Microscopically, these are spin-triplet electron-hole pairs. More generally, excitations comprised of finitely many electrons can either have odd charge and half-integer spin or even charge and integer spin. Violations of this rule imply the fractionalization of electrons, e.g., a splitting between their spin and charge. Experimental indications of this phenomenon arise, e.g., in high-temperature superconductors\cite{Lee2006,Keimer2015} and frustrated magnets.\cite{Savary+Balents2016,organicsreview,Balents2010,Zhou2017,qslbroholm}

Excitations with fractional quantum numbers cannot be created locally or captured with perturbative methods. To describe them efficiently, `dual' formulations, which encode microscopic degrees of freedom non-locally, have proven very powerful. A well-known example is the boson-vortex duality in two dimensions.\cite{Peskin1978,Dasgupta1981} The vortices are the elementary topological excitations of a bosonic superfluid but can be introduced without reference to any specific state. A vortex operator at $\vect{r}$ is defined to yield a phase $\alpha_{\vect{r},\vect{r}'}$, with $\oint\vect \nabla \alpha\cdot d\vect s=2\pi$, when interchanged with a boson operator at $\vect{r}'$. For this relation to be satisfied, the vortex must be non-local in terms of the microscopic bosons. Conversely, a microscopic boson is represented non-locally in terms of vortices, i.e., it is a `vortex in the vortex.'

The utility of this duality for capturing fractionalization follows directly from the conjugate nature of bosons and vortices. Periodicity of the one implies a specific quantization of the other, similar to position and momentum. For example, condensation of vortices with a flux of $4\pi$ (in units of $\hbar c/e$) but not the elementary $2\pi$ ones\cite{balents1999,Balents2000} results in a phase with half-charge excitations. Systems describable as bosons at half-filling may exhibit this type of vortex condensation. The average boson number translates to a $\pi$-flux background for the vortices, which leads to their band structure exhibiting two degenerate valleys. We thus expand each lattice vortex into two `low-energy' vortices as $\hat V_{\vect R} \sim \hat v_{1,\vect R}\psi_{1,\vect R}+\hat v_{2,\vect R}\psi_{2,\vect R}$, where $\psi_i$ are linear combinations of vortex wave functions at the two minima. Condensing the vortex pair $v_1 v_2$ but not individual vortices may become energetically favorable, e.g., due to their transformations under lattice symmetries.

This perspective on fractionalization has been explored in quantum magnets\cite{Sachdev2002,SVBSF2004,SBSVF2004} and superconductors\cite{Lannert2001} (see also Ref.~\onlinecite{jiang2019} for a one-dimensional analogue). In the former case, the spins are represented through hard-core bosons, which are at half-filling for vanishing magnetization. The condition for two low-energy vortex flavors is thus satisfied. Phases resulting from $v_1 v_2$ condensation exhibit fractionalization in the spin sector, i.e., they host spin-1/2 neutral excitations dubbed spinons. In superconductors, spin-singlet Cooper pairs play the role of the microscopic bosons. The condition for two low-energy vortex flavors is met at electronic half-filling. In this case, fractionalization occurs in the charge sector instead; the phases with $\langle v_1 v_2 \rangle \neq 0$ host charge-$e/2$ excitations.

In this work, we apply a similar approach to \textit{both} the spin and charge sectors of two-dimensional bosonic or electronic systems. Specifically, we insist that total charge and one spin component are conserved, i.e., we study models with $U(1)\times U(1)$. A larger $SU(2)$ spin-rotation symmetry may still be present but will not be kept manifest in our dual description. Its status is thus similar to the case of Abelian bosonization in one dimension.\cite{GiamarchiBook2004}  We find that at low energies, the dual theory is comprised of three vortices. The topological defects therein encode spinons and chargons -- fractional quasiparticles that carry the spin and charge, respectively, of the microscopic electrons or bosons. 

The rest of the paper is organized as follows: In Sec.~\ref{sec.SCS_duality}, we introduce the spin-charge separation duality in three steps. We begin with a discussion of topological defects and their symmetries. Next, we formalize these considerations to derive a field-theoretic duality. Finally, we provide a concrete implementation thereof within the coupled-wire framework. Two generalizations of the duality are then provided in Sec.~\ref{sec.generalize}. In Sec.~\ref{sec.phasesandtransitions}, we describe various symmetry-broken phases and their transitions within the dual formulation. Sec.~\ref{sec.parent} applies the coupled-wire implementation of the duality to derive parent Hamiltonians for phases that host fractional excitations. The paper concludes in Sec.~\ref{discussion} with a discussion of possible applications and extensions.

\section{Spin-charge separation (SCS) duality}
\label{sec.SCS_duality}
We first formulate the duality for two conserved flavors of bosons $B_\sigma$ at half-filling. Initially, we treat each flavor independently; interactions between different flavors will eventually be included based on symmetries. The analysis summarized in the introduction applies to each flavor separately, and the dual formulation of the model contains four low-energy vortices $v_{\sigma,l}$. Their numbers are not separately conserved. Specifically, the operators $\hat M_\sigma \equiv v_{\sigma,1}^\dagger v_{\sigma,2}$ have no net vorticity and carry neither charge nor spin. As such, their presence in any effective low-energy theory can only be excluded based on discrete symmetries. For example, $M_\sigma$ must be odd under lattice translations. Otherwise, there would be a single minimum in the vortex band structure, despite the $\pi$ flux background. Conversely, breaking translation symmetry induces unit filling in a doubled unit cell. Adding the operators $M_\sigma$ to the half-filled theory yields a single species of low-energy vortices. 

Our interest lies in the translation-symmetric case. We parametrize the vortex-mixing terms via $\hat M_c \equiv M_\uparrow M_\downarrow$ and $ \hat M_s = M_\uparrow M_\downarrow^\dagger $. Both operators are translation-symmetric yet transform non-trivially under other discrete symmetries, e.g., lattice rotations. Indeed, under square lattice symmetries, $\hat M_s$ and $\hat M_c^2$ are invariant, but $\hat M_c$ is not.\cite{Lannert2001} The higher symmetry of $\hat M_s$ compared to $\hat M_c$ arises from a cancellation between $M_{\uparrow,\downarrow}$. Thus, vortex mixing is described by the Hamiltonian density
\begin{align}
\label{eq.H_M}
 H_M \sim \lambda \hat M_s^{n_s} + \kappa \hat M_c^{n_c}+\text{H.c.}~,
\end{align}
with $n_s \leq n_c$. Our goal is to describe spin-charge separation without committing to a specific set of symmetries, which leaves $n_{c,s}$ unspecified. 

\begin{figure}[tb]
\centering
 \includegraphics[width=\columnwidth]{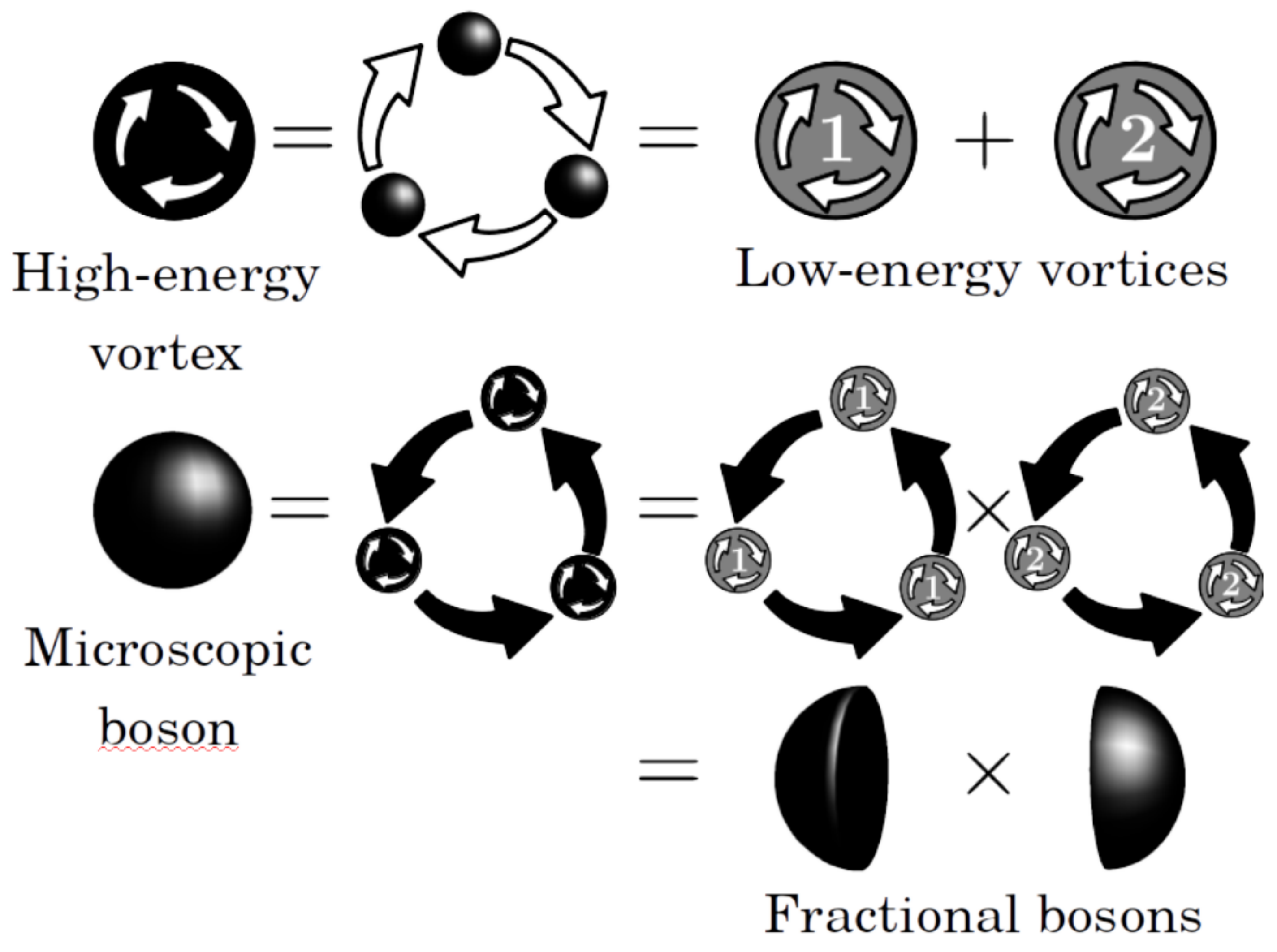}
 \caption{Dual perspective on fractionalization of bosons at half-filling. First line: a lattice-scale vortex in the microscopic boson field is expanded into two low-energy vortices. Second line: a microscopic boson is represented as the elementary defect in the lattice-scale vortex and thus in both low-energy vortices. Defects in an individual low-energy vortex describe fractional excitations. \label{fig.duality}}
\end{figure} 

The topological defects $ d_{\sigma,l}$ in $v_{\sigma,l}$ encode the quasiparticles of the theory (see Fig.~\ref{fig.duality}). The microscopic bosons are represented as $2\pi$ defects in \textit{both} low-energy vortices of matching spin, i.e., $B_\sigma = d_{\sigma,1} d_{\sigma,2}$. Single $d_{\sigma,l}$ describe fractional excitations and do not carry well-defined physical quantum numbers that are independent of a specific ground state. Still, it is beneficial to associate them with half the quantum numbers of $B_\sigma$, i.e., charge $e/2$ and spin $\pm 1/4$. In particular, this perspective suggests introducing charge-neutral spin-$\frac{1}{2}$ spinons as $b_\sigma=d_{\sigma,2}d^\dagger_{-\sigma,1}$. Crucially, the spinons do not acquire any non-trivial phases upon encircling $\hat M_s$.

The complete set of low-energy defects may be encoded through $b_{\sigma}$ and $d_{\sigma,1}$. Equivalently, we parametrize low-energy vortices by the duals of $b_{\sigma}$ and $d_{\sigma,1}$, i.e., $v_{b_\sigma} \equiv v_{\sigma,2}$ and $v_{d_\sigma} \equiv v_{\sigma,1}v_{-\sigma,2}$. In terms of the new vortex variables, the operators $M_{c,s}$ take the form
\begin{subequations}
\label{Eq.M_cs_bd}
\begin{align}
 \label{Eq.M_cs_bd-c}
 \hat M_c &=v_{d_\uparrow}^\dagger v_{d_\downarrow}^\dagger v_{b_\uparrow}^2v_{b_\downarrow}^2~,\\
 \label{Eq.M_cs_bd-s}
 \hat M_s &=v_{d_\uparrow}^\dagger v_{d_\downarrow}~.
\end{align}
\end{subequations}
When individual $\hat M_s$ are permitted ($n_s=1$ in $H_M$), the two flavors $v_{d_\sigma}$ hybridize, and their band minima split. We discard the one at higher energies and replace $v_{d_\sigma}$ with the linear combination $v_c \sim\sum_\sigma v_{d_\sigma}$. Its dual, $c\equiv d_{\uparrow,1} d_{\downarrow,1}$, is a spinless charge-$e$ `chargon.' As the elementary defect in a single vortex, it is at integer filling. Following the same logic, we conclude that also the spinons are at integer filling. We address this apparent paradox in Sec.~\ref{connectiontopartons} and within the coupled-wire realization of Sec.~\ref{sec.wirerealization}. The various quasiparticles are summarized in Tab.~\ref{tab.tab}.

\begin{table}[tb]
 \centering
 \caption{Representation of various quasparticles through defects in the low-energy vortices and their (apparent) quantum numbers.}
 \begin{tabular*}{\linewidth}{@{\extracolsep{\fill}}c|rrrr|cr}
 \hline \hline 
 Quasiparticle& \multicolumn{4}{c|}{Defects in} & Charge&Spin\\
 &$v_{\uparrow,1}$ &$v_{\uparrow,2}$&$v_{\downarrow,1}$ & $v_{\downarrow,2}$&& \\\hline
Boson $B_\uparrow$ & $ 2\pi$ & $ 2\pi$ & & & $1$ & $\hfill 1/2$\\
Boson $B_\downarrow$ & & & $ 2\pi$ & $ 2\pi$ & $1$ & $-1/2$\\
Spinon $b_\uparrow$ & & $ 2\pi$ &  $-2\pi$ & & 0& $\hfill 1/2$ \\
Spinon $b_\downarrow$&  $-2\pi$ & & & $ 2\pi$ &0 & $-1/2$ \\
Chargon $c$ &  $ 2\pi$ & &  $ 2\pi$ & & $1$ &  0\hphantom{$2$}\\\hline \hline 
 \end{tabular*}
 \label{tab.tab}
\end{table}

\subsection{Field-theoretic derivation}
\label{sec.FT_der}
To formally implement the duality, we describe the microscopic lattice bosons with the (Euclidean) action
\begin{align}
\label{eq.bosons}
 \nonumber {\cal S}_\text{Boson} =& \int_\tau\sum_{\vecta r,\sigma} B^*_{\vecta r,\sigma}(\partial_\tau -iA^\sigma_{\tau,\vecta{r}})B_{\vecta r,\sigma} \\
 & + \int_\tau \sum_{\vecta r,\vecta{r'},\sigma }t_{\vecta r,\vecta{r'}} B^*_{\vecta r,\sigma}e^{i A^\sigma_{\vecta r,\vecta r'}}B_{\vecta{r'},\sigma}+\ldots
\end{align}
Here $A^\sigma \equiv A^c+\sigma A ^s$ with $\sigma=+,-$ for spin $\uparrow,\downarrow$. $A^c$ is the electromagnetic vector potential, and $A^s$ is a `spin' vector potential, which we include to keep track of the spin degree of freedom through the manipulations below. The ellipses denote symmetry-allowed short-range interactions within and between species; we insist on translation symmetries and conservation of spin and charge. Additional discrete symmetries, such as lattice rotations, may play an important role and will also be discussed when they do.

The standard boson-vortex duality expresses the conserved three currents of the two species as field strengths of emergent non-compact gauge fields $\tilde a^\sigma$. The latter thus couple to the external probing fields $A^\sigma$ via the Lagrangian density
\begin{align}
 \mathcal{L}_{Ada} & = \tfrac{i}{2\pi} \sum\nolimits_\sigma A^\sigma d \tilde a^\sigma ~.
 \label{eq.low_e_vortices-adA}
\end{align}
Vortex configurations in $B_\sigma$ are encoded by vortices $V_\sigma$ that reside on dual lattice sites $\vecta{\tilde{r}}$.\cite{Peskin1978,Dasgupta1981} The model ${\cal L}_\text{Boson}$ is then expressed through the lattice vortex action ${\cal S}_\text{Vortex}={\cal S}_V- \mathcal{S}_{Ada}$ with
\begin{align}
\label{eq.lattice_vortices}
 \nonumber {\cal S}_{V} =& \int_\tau \sum_{\vecta{\tilde r} ,\sigma} V^*_{\vecta {\tilde r},\sigma}(\partial_\tau -i\tilde{a}^\sigma_{\tau,\vecta{\tilde r}})V_{\vecta {\tilde r},\sigma}\\
 & + \int_\tau \sum_{\vecta {\tilde r},\vecta {\tilde r}',\sigma}\tilde t_{\vecta {\tilde r},\vecta {\tilde r}'} V^*_{\vecta {\tilde r},\sigma}e^{i \tilde a^\sigma_{\vecta {\tilde r},\vecta {\tilde r}'}}V_{\vecta {\tilde r}',\sigma}+\ldots
\end{align}
Here, the ellipses include symmetry-allowed interactions for the vortices and kinetic terms for the gauge fields. The latter arise with non-universal coefficients of order unity.

Half-filling of the microscopic bosons translates to a background flux of $\langle \vecta{\Delta} \times \vecta a^\sigma\rangle=\pi$ per dual-lattice plaquette. We choose a gauge for this flux, which requires (at minimum) a two-site unit cell. Consequently, there are two energy bands per vortex species. The lower band exhibits two degenerate minima in the reduced Brillouin zone, reflecting that physical symmetries are not broken by gauge choices. We expand each vortex $V_\sigma$ around its minima and discard the upper band. These steps yield four low-energy vortices $v_{\sigma,l}$, which transform non-trivially under lattice symmetries (see Refs.~\onlinecite{Lannert2001,Sachdev2002}).

By discarding the upper bands, we have eliminated half of the degrees of freedom. Consequently, $v_{\sigma,l}$ (unlike $V_\sigma$) provide a dual description for ${\cal L}_\text{Boson}$ only near half-filling and at low energies. We describe them by the continuum Lagrangian density ${\cal L}_v = \sum_{l}\mathcal{L}_{l}- \mathcal{L}_{adA}$ with 
\begin{align}
\label{eq.low_e_vortices}
 \mathcal{L}_{l} & =\sum_\sigma \big[|(\vectb\nabla - i \vectb{\tilde{a}}^\sigma)v_{\sigma,l}|^2-m|v_{\sigma,l}|^2+\ldots\big]~.
\end{align}
Here, the ellipsis includes kinetic terms for the gauge fields $\tilde a^\sigma$, symmetry-allowed interactions such as $|v_{\sigma,l}|^{4}$, and the vortex-mixing terms of Eq.~\eqref{eq.H_M}. 
 
We proceed by rewriting the vortex theory in terms of $v_{b_\sigma}$ and $v_{d_\sigma}$ defined in Sec.~\ref{sec.SCS_duality}. The former are identical to $v_{\sigma,2}$ and are thus described by $\mathcal{L}_{2}$ of Eq.~\eqref{eq.low_e_vortices}. The latter couple to both $\tilde a^{\uparrow},\tilde a^{\downarrow}$ with unit charge. They are thus governed by
\begin{align}
 \label{eq.low_e_vortices-d}
 \nonumber \mathcal{L}_{d} =& \sum_\sigma \big[ |(\vectb\nabla - i \vectb{\tilde{a}}^c)v_{d_\sigma}|^2-m_d|v_{d_\sigma}|^2\big]\\
 &+\big[ \lambda (v_{d_\uparrow}^* v_{d_\downarrow})^{n_s}+\text{H.c.}\big]+\ldots ~, 
\end{align}
with $\tilde a^c=\tilde a^\up+\tilde a^\down$. We have explicitly included the first term of Eq.~\eqref{eq.H_M} to emphasize its key role here. In particular, for $n_s=1$, the second line in $\mathcal{L}_d$ amounts to substituting $v_{d_\sigma}\rightarrow v_c$ in the first line. This value of $n_s$ is consistent, e.g., with translations and rotations on a square lattice. Indeed, defining $v_{\sigma,l}$ as in Ref.~\onlinecite{Lannert2001}, we find that $v_{d_\uparrow}$ and $v_{d_\downarrow}$ interchange under $x$- or $y$-translations and are invariant under ${C_4}$ rotations. 

We focus on $n_s=1$ and perform a second set of boson-vortex dualities. Specifically, we dualize each of the three low-energy variables $v_c,v_{b_\uparrow},v_{b_\downarrow}$ separately into $c,b_{\uparrow},b_{\downarrow}$ and three new emergent gauge fields $a,a_\uparrow,a_\downarrow$. 
Subsequently, integrating over $\tilde a^\sigma$ enforces the two constraints $\vectb{a}^\sigma = \vectb{A}^\sigma -\vectb{a}$.
We arrive at $\mathcal{L}_\text{Boson}^\text{SCS} = \mathcal{L}_{c} + \mathcal{L}_b + \mathcal{L}_a$ with
\begin{subequations}
\label{eq.bosonic_scs}
\begin{align}
 \label{eq.bosonic_scs-c}
 \mathcal{L}_{c} &=|(\vectb \nabla - i \vectb a - i \vectb A^c)c|^2+\ldots~,~
 \\
 \label{eq.bosonic_scs-b}
 \mathcal{L}_{b} &= \sum_\sigma |(\vectb \nabla + i\vectb a -i\sigma \vectb A^s) b_\sigma|^2+\ldots~, 
 \\
 \label{eq.bosonic_scs-a}
 \mathcal{L}_a &=\kappa M_c^{n_c}+\text{H.c.}+\ldots~,
\end{align}
\end{subequations}
where the ellipsis includes a Maxwell term for $a$ with a coefficient of order unity. The $\kappa$-term introduces $4\pi n_c$ monopoles into the gauge field. The latter is thus compact, in contrast to $\tilde a^\sigma$. Eq.~\eqref{eq.bosonic_scs} is a key result of our work. It describes a representation of microscopic bosons as $B_\sigma =b_\sigma c$. Crucially, deriving it did not require us to insist on such a decomposition on the lattice scale nor invoke a specific saddle point. Instead, it arose naturally from the low-energy properties of vortices that encode bosons at half-filling. 

\subsubsection{Connection to partons}
\label{connectiontopartons}
The Lagrangian in Eq.~\eqref{eq.bosonic_scs} strongly resembles the outcome of parton approaches.\cite{Lee2005,Lee2006} Specifically, the microscopic boson operator may be expressed as
\begin{align}
 B_{\sigma,\vecta r}= \tilde b_{\sigma,\vecta r} \tilde c_{\vecta r}~,
\end{align} 
with the constraint $\tilde c_{\vecta r}^\dagger \tilde c_{\vecta r} = \sum_{\sigma}\tilde b_{\sigma,\vecta r}^\dagger \tilde b_{\sigma,\vecta r}$. This representation is suitable for exploring saddle points where spinons $\tilde b$ and chargons $\tilde c$ decouple, i.e.,
\begin{align}
 B_{\sigma,\vecta r}^\dagger B_{\sigma,\vecta r'} \xrightarrow{\text{mean-field}} \chi_{\tilde b_\sigma} \tilde c_{\vecta r}^\dagger \tilde c_{\vecta r'} +\chi_{\tilde c} \tilde b_{\sigma,\vecta r}^\dagger \tilde b_{\sigma,\vecta r'}~,
\end{align}
with $\chi_{\tilde b_\sigma}=\langle \tilde b_{\sigma,\vecta r}^\dagger \tilde b_{\sigma,\vecta r'}\rangle$ and $\chi_{\tilde c}=\langle\tilde c_{\vecta r}^\dagger \tilde c_{\vecta r'} \rangle$. Phase fluctuations of the mean-field parameters $\chi_{\tilde c,\tilde b}$ take the form of an emergent $U(1)$ gauge field. The low-energy field theory for $\tilde c,\tilde b$ is thus of the form given by Eq.~\eqref{eq.bosonic_scs}.

The density of $\tilde{b}_\sigma$ spinons is identical to that of $B_\sigma$. In particular, at half half-filling, vortices in either exhibit two minima in the band structure. By contrast, the spinons $b_\sigma$ introduced above are each dual to a single low-energy defect, indicating that they are at integer filling. Still, the total number of $b_\sigma$ spinons and $B_\sigma$ agree. The resolution to this apparent paradox lies in the steps between Eqs.~\eqref{eq.lattice_vortices} and \eqref{eq.low_e_vortices}. Discarding the high-energy band amounts to working with a doubled unit cell, for which the microscopic bosons are at unit filling. Crucially, we have not committed to any specific choice of unit cell at this stage and thus did not break any symmetries. However, the lattice symmetries are no longer manifest. They are broken, e.g., by the seemingly-innocuous formation of a Mott state with one spinon $b_\sigma$ per unit cell.

The partons $\tilde b$ permit more straightforward access to some phases, e.g., ones that break lattice symmetries. On the other hand, the formulation in terms of $b$ allows more direct access to unconventional phase transitions (cf.~Sec.~\ref{sec.phasesandtransitions}). Finally, the dual field theory has a gauge-field coupling of order unity, as in the conventional boson-vortex duality.~\cite{Peskin1978,Dasgupta1981} By contrast, the gauge-field coupling in parton-based derivations is formally infinite at the lattice scale. Approximate methods such as variational techniques may thus be more reliable in the dual formulation.

\subsubsection{Fermions}
\label{sec.fermionscs}
A fermionic version of Eq.~\eqref{eq.bosonic_scs} is readily generated via flux attachment.\cite{Wilczek1982,Fradkin1989} There are multiple ways to achieve this, and the most suitable choice depends on the system in question. In the quantum Hall context, statistical flux quanta are typically bound to the electric charge.\cite{Jain_composite_2007} Alternatively, Ref.~\onlinecite{Balents2000} considered flux attachment to the spin and described various superconductors. Either of these choices yields a fermionic analogue of Eq.~\eqref{eq.bosonic_scs}. Here, we adopt a different prescription that closely aligns with the spirit of the derivation above. 

We introduce independent Chern-Simons gauge fields for each species and attach fluxes with opposite signs. Formally, we construct the fermionic theory as
\begin{align}\label{fluxattach}
{\cal L}_\text{Fermion} = {\cal L}_\text{Bosons}\big|_{A^\sigma \rightarrow A^\sigma + a^\sigma_\text{CS}}+ {\cal L}_{\text{CS}}~,
\end{align}
where the statistical gauge fields $a^\sigma_\text{CS}$ are governed by 
\begin{align} \label{eq.LCS}
 {\cal L}_{\text{CS}}=i\sum_{\sigma} \frac{\sigma}{4\pi} a_{\text{CS}}^\sigma d a_{\text{CS}}^\sigma~.
\end{align}
Notice that ${\cal L}_{\text{Fermion}}$ is symmetric under time reversal, with $\mathcal{T} : a_{\text{CS}}^{\sigma} \rightarrow a_{\text{CS}}^{-\sigma}$. The presence of ${\cal L}_{\text{CS}}$ does not affect the manipulations performed above. To obtain the fermionic duality, we replace $ A^\sigma \rightarrow A^\sigma + a^\sigma_\text{CS}$ in Eq.~\eqref{eq.bosonic_scs} and add ${\cal L}_{\text{CS}}$ to both sides. Shifting the gauge field according to $a \rightarrow a+ \textfrac{1}{2}(a^\uparrow_\text{CS}+a^\downarrow_\text{CS})$ decouples the chargon from the Chern-Simons gauge fields and changes the spinon Lagrangian to
\begin{align}
 \mathcal{L}'_{b} &= \sum_\sigma |(\vectb \nabla + i\vectb a+ ia^\sigma_{\text{CS},\mu} -i\sigma \vectb A^s) b_\sigma|^2+\ldots~
\end{align}
Finally, the model $\mathcal{L}'_{b}+ {\cal L}_{\text{CS}}$ describes fermionic spinons minimally coupled to the dynamical gauge field $a$ and the probing field $\sigma A^s$. 

We have thus achieved a representation of microscopic fermions $F_\sigma$ in terms of a bosonic chargon and fermionic spinons as $F_\sigma = f_\sigma c$. We point out that the spinons' integer filling (cf.~Sec.~\ref{connectiontopartons}) implies that metallic states with uncompensated particle or hole pockets are beyond the dual low-energy theory.

\subsubsection{Doping}
We derived the spin-charge separation duality specifically for the case of half-filling. However, the presence of $ A_\sigma$ on either side suggests that they remain operable after doping the system. Indeed, small deviations of the filling amount to an additional flux $2\pi \delta n$ per plaquette for the lattice vortices. For $\delta n \ll 1$, the additional flux can be incorporated perturbatively after introducing the low-energy vortices. The flux experienced by the latter acts as a chemical potential for the chargons and spinons, modifying their density accordingly. 

\subsection{Coupled-wire realization}
\label{sec.wirerealization}
To implement the schematic manipulations on a concrete model, we use the `coupled-wire' framework. There, duality can be established by an explicit, non-local mapping between quantum partition functions. All the ingredients required for deriving the SCS-duality are readily available: Refs.~\onlinecite{Mross2016,Mross2017} implemented boson-vortex duality and flux attachment. Using these results, retracing the steps leading to Eq.~\eqref{eq.bosonic_scs} and its fermionic counterpart is straightforward. We derive the mapping in Appendix \ref{App.A} and summarize only the final result here. 

We consider an array of one-dimensional wires extending along the $x$-direction and labeled by integers $y$. Each wire hosts electrons close to half-filling, i.e., with $k_F \approx \frac{\pi}{2}$, and is described by a Luttinger-liquid Hamiltonian. The long-wavelength expansion of the electron operators is
\begin{align}
 \psi_{y,\sigma}(x) = e^{i k_F x} \psi_{y,\sigma,R}(x) + e^{-i k_F x} \psi_{y,\sigma,L}(x)+\ldots
\end{align}
Each species of fermionic spinons resides on half the dual wires $\tilde{y}\equiv y+1/2$.\cite{Leviatan2020} We expand them analogously as
\begin{align}
 f_{\tilde{y},\sigma}(x) = e^{i 2k_F x} f_{\tilde{y},\sigma,R}(x) + e^{-i 2 k_F x} f_{\tilde{y},\sigma,L}(x)+\ldots~,
\end{align}
with $\sigma=\up$ for odd $\tilde{y}$ and $\sigma=\down$ for even $\tilde{y}$. By contrast, the bosonic chargons occupy all wires. They are expanded as
\begin{align}
 \tilde c_y(x) = c_y(x) [1 + e^{ i 4k_F x } P_y(x) + \text{H.c.}] + \ldots ~,
\end{align}
where $P_y(x)$ creates a $2\pi$ phase slip in $c_y$ at position $x$. We will suppress the $x$-dependence of operators and the wire index when unneeded from here on.

Individual spinons and chargons are non-local. This aspect is reflected in their intra-wire kinetic terms, which take the form of Luttinger liquids minimally coupled to an emergent photon.\cite{Mross2016,Mross2017,Fuji2019,Leviatan2020} Nevertheless, many of their basic processes are locally expressible. For the chargons, we find\footnote{the chirality on the r.h.s. of Eq.~\eqref{eq.WA-F.Der.psi-c-terms-2} by the choice of the overall sign in Eq.~\eqref{fluxattach}}
\begin{subequations}
\begin{align}
\label{eq.WA-F.Der.psi-c-terms-1}
 P_{y}&= \psi^\dagger_{y,\up,R}\psi_{y,\up,L}\psi^\dagger_{y,\down,R}\psi_{y,\down,L} ~,\\
 \label{eq.WA-F.Der.psi-c-terms-2}
 c^\dagger_{y+1}c_{y}&=
 \begin{cases}
 \psi^\dagger_{y+1,\down,L}\psi_{y,\down,R}\quad \text{$y$ even}~,\\
 \psi^\dagger_{y+1,\up,R}\psi_{y,\up,L}\quad \text{$y$ odd}~.
 \end{cases}
\end{align}
\end{subequations}
In particular, a Mott state of bosonic chargons, driven by the proliferation of phase slips $P_y$, corresponds to a Mott insulator of microscopic electrons on each wire. 

Spinon inter-wire hopping is conveniently expressed in terms of the microscopic spin-densities near $\pm 2k_F$, i.e., $ \vec{s}_{R} = \psi^\dagger_{\tau,R}\vec{\sigma}_{\tau,\tau'}\psi_{\tau',L}$ and $ \vec{s}_{L} = \psi^\dagger_{\tau,L}\vec{\sigma}_{\tau,\tau'}\psi_{\tau',R}$, as
\begin{subequations}
\label{spinonhop}
\begin{align}
 f^\dagger_{2\tilde{y}+1,\up,R}f_{2\tilde{y}-1,\up,L}&= s^+_{2y+1,R} s^-_{2y,L} ~,\\
 f^\dagger_{2\tilde{y}-2,\down,R}f_{2\tilde{y},\down,L}&= s^-_{2y,L} s^+_{2y-1,R}~. 
\end{align}
\end{subequations}
 Notice that $\vec s_R = P \vec s_L$. Consequently, in the Mott state where $\langle P \rangle \neq 0$, the two are identical and reduce to the N\'eel vector. Intra-wire spinon umklapp processes translate to
\begin{align}
 f^\dagger_{\tilde{y},\sigma,R}f_{\tilde{y},\sigma,L}= \psi^\dagger_{y+1,\sigma,R}\psi_{y+1,\sigma,L} \psi^\dagger_{y,\sigma,R}\psi_{y,\sigma,L}~,
\end{align}
with $\sigma=\downarrow$ for even $\tilde{y}$ and $\sigma=\uparrow$ for odd. These, along with the inter-wire hopping, facilitate a local description for any spinon band structure. 

Finally, certain microscopic single-fermion operators have a simple expression in terms of $c$ and $f$. We find
\begin{subequations}
\label{eq.WA-F.Der.psi-c-f}
\begin{align}
 \psi_{2y,\sigma,L} &= c_{2y} f_{2y-\tfrac{\sigma}{2},\sigma,L}~,\\
 \psi_{2y+1,\sigma,R} &= c_{2y+1} f_{2y+\tfrac{\sigma}{2},\sigma,R}~.
\end{align}
\end{subequations}
These microscopic fermions become identified with their spinon counterparts once chargons are condensed. We elaborate on this situation and other sample phases for the chargons in Sec.~\ref{sec.parent}.

\section{Generalizations}
\label{sec.generalize}
The derivation in Sec.~\ref{sec.SCS_duality} follows three readily-generalizable steps: (i) Duality is performed on the high-energy degrees of freedom. (ii) The low-energy limit is taken within the dual description. (iii) Another duality is applied to the long-wavelength continuum variables. We now describe two additional examples of this approach.

\subsection{Interchanging spin and charge}
In the duality of Sec.~\ref{sec.SCS_duality}, we describe the microscopic bosons in terms of one chargon and two spinons. The structure of low-energy defects $d_{\sigma,l}$ suggests an alternative description in terms of a single spinon $s$ and two chargons $h_{\pm}$. To construct them, we parametrize the low-energy vortices by
\begin{equation}
\begin{split}
v_+ &= v_{\uparrow,1}~, \qquad \qquad
v_\up = v_{\down,1} v_{\up,1}^\dagger~,\\
v_- &= v_{\downarrow,2}^\dagger~, \qquad \qquad 
v_\down = v_{\up,2} v_{\down,2}^\dagger~.
\end{split}
\end{equation}
The chargons $h_+ = d_{\up,1} d_{\down,1}$ and $ h_- = d_{\up,2}^\dagger d_{\down,2}^\dagger$ are dual to $v_\pm$. Moreover, $M_s=v_{\downarrow}^\dagger v_\uparrow$, so for $n_s=1$ the low energy vortices $v_\sigma$ hybridize into a single vortex $v_s\sim v_\up+v_\down^\dagger$ whose dual is $s = d^\dagger_{\downarrow,1}d_{\uparrow,2}$. 

The field-theoretic derivation proceeds as before, and we arrive at $\mathcal{L}_\text{Boson}^\text{CSS} = \mathcal{L}_{h} + \mathcal{L}_s + \mathcal{L}_a$ with
\begin{subequations}
\label{eq.bosonic_CSS}
\begin{align}
\label{eq.bosonic_CSS-s}
 \mathcal{L}_{s} &= |(\vectb \nabla - i \vectb a - i \vectb A^s)s|^2+\ldots~,~
 \\
\label{eq.bosonic_CSS-h}
 \mathcal{L}_{h} &= \sum_{\tau=\pm} |(\vectb \nabla + i\vectb a -i\tau \vectb A^c) h_\tau|^2+\ldots~,
\end{align}
\end{subequations}
and $\mathcal{L}_a$ as in Eq.~\eqref{eq.bosonic_scs-a}. Here, microscopic bosons are represented as $B_\up = h_+ s$ and $B_\down = h_-^\dagger s^\dagger$. This charge-spin separation (CSS) is complementary to that of Sec.~\ref{sec.SCS_duality}, and either may be preferable, depending on the specific question. As before, a fermionic representation $F_\up = \psi_+ s$ and $F_\down = \psi_-^\dagger s^\dagger$ with a bosonic spinon $s$ and fermionic chargons $\psi_\pm$ is obtained via flux attachment. 

\subsection{$N$ flavors of bosons}
So far, we considered the most prevalent physical setup, i.e., two flavors at half-filling. However, the SCS-duality is readily adapted for $N\geq2$ flavors of microscopic bosons, all at filling $\nu=\tfrac{1}{N}$. Here, we describe the generalized SCS-duality and focus on the characterization of topological defects at low energies. The generalized duality reduces to the one in Sec.~\ref{sec.SCS_duality} for $N=2$.

The lattice vortices dual to each flavor $\alpha=1,\ldots N$ of the $\tfrac{1}{N}$-filled microscopic bosons experience a background flux of $\tfrac{2\pi}{N}$. Their band structures thus exhibit $N$ degenerate valleys in the lowest energy band. We label these minima by $l=1,\ldots N$ and expand each flavor of lattice vortices. The dual description is thus comprised of $N^2$ low-energy vortices. We describe them by $v_{\alpha,l}$, defined such that $\hat x$-translations cyclically permute them, $l\rightarrow l+1 \mod N$, and $\hat y$-translations introduce $l$- but not $\alpha$-dependent phases.

With this parametrization, we construct $(N-1)^2$ vortex-mixing terms analogous to $M_s$ of Sec.~\ref{sec.SCS_duality}, i.e.,
\begin{align}
 M_{s}^{\alpha,l} = v^\dagger_{\alpha,l} v^\dagger_{\alpha+1,l+1} v_{\alpha,l+1} v_{\alpha+1,l}
\end{align}
with $l,\alpha=1,\ldots N-1$. These have no net vorticity and are invariant under $y$-translations. Under $x$-translations, they permute according to $l\rightarrow l+1 \mod N-1$. Consequently, a translation-symmetric theory permits all $M_{s}^{\alpha,l}$ to linear order. Each $M_s$ introduces hybridization between two vortices, and the number of independent low-energy vortices reduces to $2N-1$. Equivalently, the individual topological defects $d_{\alpha,l}$ in $v_{\alpha,l}$ are confined. By contrast, the composites
\begin{align}
\begin{split}
 \label{eq.spinonN}
 b_\alpha &= \prod_{j} d_{\alpha,j} \Big(\prod_\beta d_{\beta,1}\Big)^\dagger ~,\\
 c_{l} &= \prod_\beta d_{\beta,l}~,
 \end{split}
\end{align}
with $\alpha=1,\ldots N$ and $l=1,\ldots N-1$, are unaffected by the vortex hybridizations. They are thus analogous to the spinons and chargons of Sec.~\ref{sec.SCS_duality}.

Microscopic bosons can be represented as
\begin{align}
 B_\alpha=\prod_j d_{\alpha,j} = b_\alpha c_1~,
\end{align}
without reference to $c_{i \neq 1}$. It is thus appealing to attribute the microscopic charge to $c_1$ and the flavor to $b_\alpha$. The remaining $c_{i\neq 1}$ are not associated with these quantum numbers and may condense without breaking physical symmetries.\footnote{
Still, their state may affect the microscopic phase. For example, we expect that a gapped state for some of the $c_{i\neq1}$ breaks lattice symmetries.} The gauge theory describing $b_\alpha$ and $c\equiv c_1$ mirrors the $N=2$ case. We introduce a probing field $A_\alpha$ for each flavor and define $A^c = \frac{1}{N}\sum_\beta A_\beta$.
The dual theory is $\mathcal{L}_{N\text{-Boson}}^{\text{SCS}}=\mathcal{L}_{b,N}+\mathcal{L}_{c} + \mathcal{L}_{a}$, with
\begin{align}
 \label{eq.Nbosonic_scs-b}
 \mathcal{L}_{b,N} &= \sum_{\alpha} |(\vectb \nabla + ia_{\mu} - i A_{\alpha,\mu} + iA^c_{\mu}) b_\alpha|^2+\ldots~,
\end{align}
and $\mathcal{L}_{c},\mathcal{L}_{a}$ of Eq.~\eqref{eq.bosonic_scs}.

\section{Simple phases and phase transitions}
\label{sec.phasesandtransitions}
The long-wavelength description derived in Sec.~\ref{sec.SCS_duality} readily captures superfluid and insulating phases with or without magnetic order. Their analysis is relatively standard. The vortex formulation, Eqs.~\eqref{eq.low_e_vortices} and \eqref{eq.low_e_vortices-d}, closely relates to dual descriptions of quantum magnets~\cite{Sachdev2002,SVBSF2004} and superconductors.\cite{Lannert2001} The spinon-chargon formulation, Eq.~\eqref{eq.bosonic_scs}, has a structure common to the low-energy theory obtained via parton decompositions. However, as discussed in Sec.~\ref{connectiontopartons}, the two are not the same. We illustrate the differences by focusing on relatively simple phases and transitions, summarized in Fig.~\ref{fig:phases}. Specifically, we describe phases of bosons characterized by spin rotation symmetry, particle number conservation, translations, and $C_4$ rotations.

\subsection{Phases}
Spatially-homogeneous phases entirely determined by the continuous symmetries are straightforward to analyze in the vortex and the spinon-chargon formulations. Both symmetries are broken in a magnetically ordered superfluid. Here, the two spinons and the chargon each form a condensate. Two of the three associated phase fluctuations correspond to the two microscopic Goldstone modes; the third gaps out along with the emergent gauge field $a$ via the Higgs mechanism. In the dual formulation, all vortices are massive. The gauge fields $\tilde a^\sigma$ remain gapless and encode the two Goldstone modes.

When only spin-rotation symmetry is broken, a magnetically-ordered insulator forms. Such a phase arises when both spinons are condensed, and the chargon forms a Mott state. The total phase fluctuations of both spinon condensates consume the emergent gauge field $a$ and become massive. Their relative fluctuations describe a single Goldstone mode, i.e., spin waves. In the vortex formulation, $v_c$ is condensed and generates a Higgs mass for $\tilde a^c$. The vortices $v_{b_\sigma}$ are massive, and the gapless photon $\tilde a^s$ encodes the spin waves. 

The converse situation where only particle number conservation is broken arises in paired superfluid of charge-$2e$ singlets. Here, the chargon and the product of both spinons acquire non-zero expectation values, yet individual spinons are gapped. This phase is expressed most naturally within the CSS formulation of Eq.~\eqref{eq.bosonic_CSS}. There, both chargons $h_\pm$ are condensed, and the spinon $s$ forms a Mott state. This paired superfluid is thus related to the magnetic insulator via the interchange of spinons and chargons. The vortex description follows analogously: $\tilde a^s$ is swallowed by fluctuations of $v_s$ while $\tilde a^c$ remains gapless and encodes the superfluid phase fluctuations.

Finally, if both continuous symmetries remain intact, the system realizes a trivial or fractional non-magnetic insulator. In the former case, spatial symmetries must break. This fact is not readily apparent in the continuum theory of Eq.~\eqref{eq.bosonic_scs}; the spinons are at unit filling, so a trivial symmetric insulator appears possible. Recall, however, that their filling refers to an enlarged unit cell (cf.~Sec.~\ref{connectiontopartons}), whose specifics are immaterial when they are condensed. By contrast, if at least one spinon is gapped, the specific unit cell is cemented. The spatial symmetry breaking is more readily apparent in the vortex formulation. There, a Mott phase of the $\sigma$ spinons corresponds to either $v_{\sigma,2}$ or $v_{-\sigma,1}$ obtaining an expectation value. Both such condensates transform non-trivially under spatial symmetries (cf.~Sec.~\ref{sec.SCS_duality}). 

Within the vortex formulation, these non-magnetic insulators are captured by Eq.~\eqref{eq.low_e_vortices}, with $m<0$. On a square lattice, the leading interaction terms are
\begin{align}
 \nonumber \mathcal L_v' = &\sum_l \Big[ u|v_{\up,l}|^4 +u|v_{\down,l}|^4 + g_s |v_{\up,l}|^2|v_{\down,l}|^2 \Big] \\
 +& \sum_\sigma \Big[ g_1 |v_{\sigma,1}|^2|v_{\sigma,2}|^2 + g_2 |v_{\sigma,1}|^2|v_{-\sigma,2}|^2\Big]~,
\end{align}
and vortex mixing is described by Eq.~\eqref{eq.H_M} with $n_s=1$ and $n_c=2$. In Sec.~\ref{sec.SCS_duality}, we took the limit $\lambda \rightarrow \infty$ first and thereby reduced the available degrees of freedom from four to three. It is instructive also to follow a complementary approach, i.e., first analyze $\mathcal L_v'$ and then add Eq.~\eqref{eq.H_M} as a perturbation. 

\begin{figure}[tb]
\centering
 \includegraphics[width=\columnwidth]{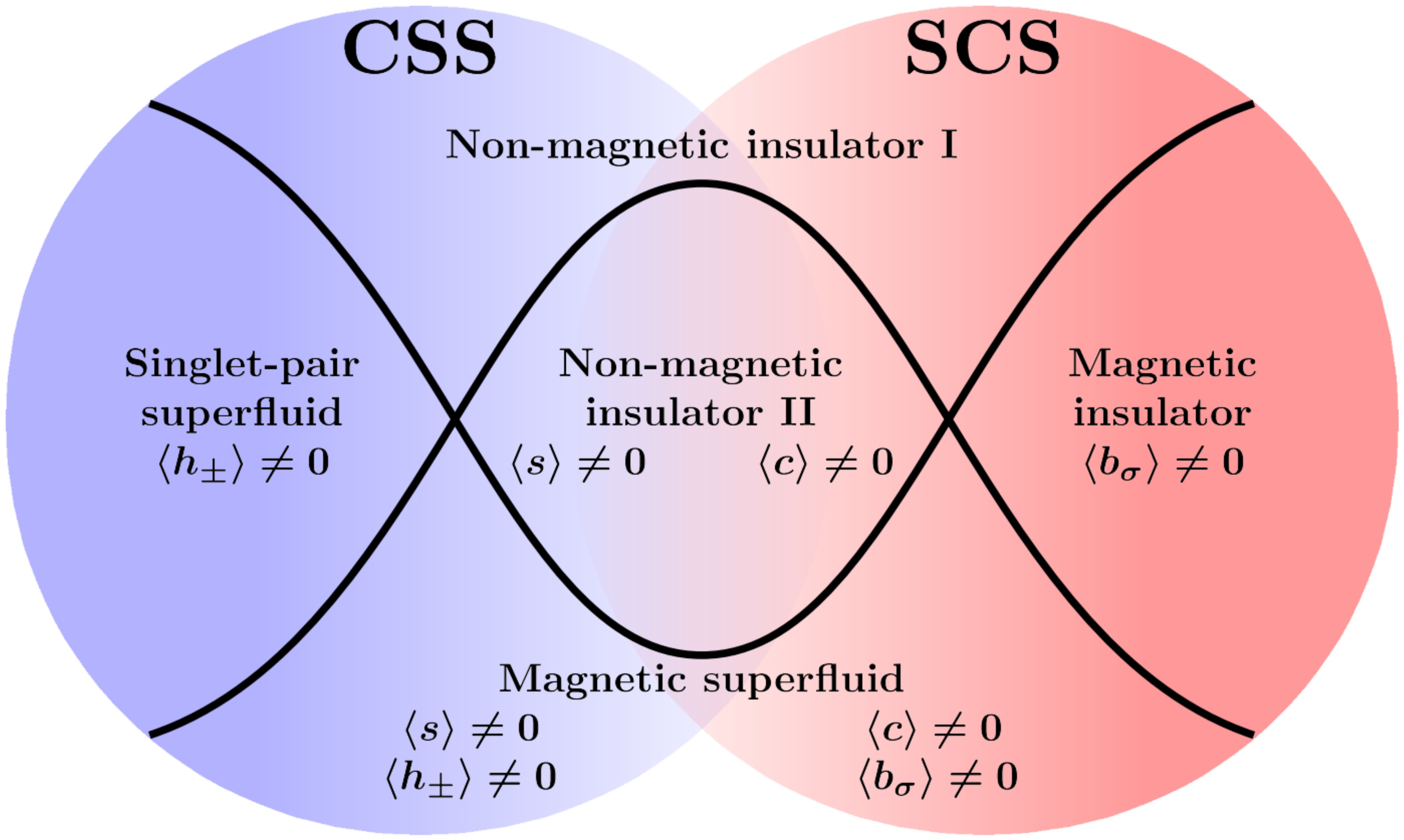}
 \caption{Phases and transitions discussed in Sec.~\ref{sec.phasesandtransitions}. The phases covered by the red (right) circle have a simple description in terms of the SCS variables. Specifically, each spinon and chargon either forms a Mott state or a condensate (as indicated in the figure). The same applies to phases covered by the blue (left) circle and CSS variables.\label{fig:phases}}
\end{figure}

For negative $g_1,g_2,g_s$, all vortices $v_{\sigma,l}$ acquire non-zero expectation values. Two of the four resulting phase fluctuations are consumed by $\tilde a^\sigma$. The remaining two are rendered massive by the terms in Eq.~\eqref{eq.H_M}. The $\kappa$ term implies four degenerate states\cite{Lannert2001,Sachdev2002}, each with a different symmetry-breaking pattern. The ground state is chosen spontaneously. In the spinon-chargon formulation, all matter fields form Mott states. The low-energy theory is described by $\mathcal{L}_a$ of Eq.~\eqref{eq.bosonic_scs-a}, with $n_c=2$. In particular, the relevant monopole contribution, $\sim M_c^2$, renders the gauge field confining and spontaneously breaks spatial symmetries. This phase is dubbed non-magnetic insulator I in Fig.~\ref{fig:phases}.

As a final example, consider the non-magnetic insulator where $v_{\sigma,2}$ condense, and $v_{\sigma,1}$ are massive, or vice versa. The two cases are related by $\hat x$-translations and may arise for large, positive $g_{1,2}$. Due to $\tilde a^\sigma$, the resulting phase is gapped even without the terms in Eq.~\eqref{eq.H_M}. In the spinon-chargon formulation of this phase, both spinons form Mott states. The two possible ground states correspond to a non-zero expectation value for $c$ or the composite $c b_\uparrow b_\downarrow$. Either carry unit charge under the gauge field $a$ and render it massive via the Higgs mechanism. This phase is dubbed non-magnetic insulator II in Fig.~\ref{fig:phases}.

Fermionic phases follow straightforwardly from the bosonic ones via the flux attachment in Eq.~\eqref{eq.LCS}. For example, Mott insulators of bosons correspond to fermionic band insulators, and a boson superfluid to an integer quantum Hall state of fermions. In particular, when both bosons are superfluid, the corresponding fermions realize a quantum spin Hall state. Finally, the paired superfluid of charge-$2e$ singlets corresponds to an $s$-wave superconductor. The linear combination $a^{\text{CS}}_\up+a^{\text{CS}}_\down$ is massive due to its coupling to the Cooper-pair condensate; $a^{\text{CS}}_\up - a^{\text{CS}}_\down$ corresponds to the Goldstone mode.

\subsection{Phase transitions}
The transitions between the phases described above are illustrated in Fig.~\ref{fig:phases}. They can be readily deduced based on symmetry considerations. For completeness, we still provide their vortex and spinon-chargon descriptions. 

We begin with transitions that do not involve any change in spatial symmetries. Such is the case, for example, for the transition between the magnetic insulator and magnetic superfluid. The two phases are distinct by charge conservation, which is spontaneously broken in the superfluid. Consequently, the transition must fall into the universality class of the classical three-dimensional XY model. The critical fluctuations decouple from the gapless spin waves, which exist on both sides of the transition due to the broken spin rotation symmetry. In the spinon-chargon formulation, both spinons are condensed throughout the transition. Their relative phase fluctuations correspond to the gapless Goldstone mode. Their total phase fluctuations are rendered massive due to the gauge field $a$, and the interactions between chargons are short-ranged. The latter then undergoes a conventional superfluid--Mott insulator transition. In the dual formulation, the vortices $v_{b_{\sigma}}$ are massive in both phases, and the photon $\tilde a^s$ describes the gapless spin waves. It decouples from $\tilde{a}^c$ and the critical fluctuations of $v_{c}$ as the latter undergoes a (dual) Mott insulator--superfluid transition.\cite{Dasgupta1981} 

Transitions determined by spatial symmetries are subtler than those discussed above. Consider, for example, non-magnetic insulators I and II of Sec.~\ref{sec.phasesandtransitions}. The combinations $v^\dagger_{\up,1}v_{\down,1}$ and $v^\dagger_{\up,2}v_{\down,2}$ are condensed on both sides of the transition due to the presence of $M_s$ in Eq.~\ref{eq.H_M}. We encode the remaining fluctuations by the $U(1)$ variables $\Phi_l = v_{l,\up}v_{l,\down}$. The transition is then governed by 
\begin{align}
 \nonumber \mathcal L_{\Phi} = & \sum_l \Big[ |(\nabla_\mu- i \tilde a^c_\mu) \Phi_l|^2 - \alpha |\Phi_l|^2+\beta |\Phi_l|^4 \Big]\\
 & + \gamma |\Phi_1|^2|\Phi_2|^2 + \kappa \Big[(\Phi_1^\dagger \Phi_2)^2+\text{H.c.}\Big]~,
\end{align}
with positive $\alpha,\beta,\gamma$. The first line of $\mathcal L_{\Phi}$ describes two $U(1)$ variables that condense independently. The second line consists of competing coupling terms. For small $\gamma$, the $\kappa$-term locks the relative phases of the two condensates and gaps the corresponding Goldstone mode. The resulting phase is non-magnetic insulator I. By contrast, for sufficiently large $\gamma$, it is preferable to have either $\Phi_{1}$ or $\Phi_{2}$ massive. Here, non-magnetic insulator II is realized. The gauge field $\tilde a^c$ is massive on both sides of the transition and thus does not directly enter the critical properties.

Perhaps the most interesting situation arises when both spatial symmetries and continuous $U(1)$ symmetries play a role at the transition. A well-known example is the transition between an easy-plane antiferromagnet and a valence bond solid.\cite{SVBSF2004,SBSVF2004} In Fig.~\ref{fig:phases} these phases correspond to the magnetic insulator and non-magnetic insulator I. In both phases, $c$ is trivially gapped and we recover the known critical theory for this transition, i.e., $\mathcal{L}_{\text{deconfined}} = \mathcal{L}_b+\mathcal{L}_a$, with $\mathcal{L}_b,\mathcal{L}_a$ of Eq.~\eqref{eq.bosonic_scs}. The same critical theory also describes the transition between the non-magnetic insulator I and the paired superfluid. This result is most apparent using the CSS-duality, i.e., by replacing $c\rightarrow s$, $b_\sigma\rightarrow h_\pm$. Here, the spinon $s$ is in a Mott state, and the chargons $h_\pm$ are deconfined at the transition.

\section{Parent Hamiltonians of fractional phases}
\label{sec.parent}
The operator mappings summarized in Sec.~\ref{sec.wirerealization} readily generate coupled-wire models for a wide variety of phases. \textit{Any} local model where spinons and chargon are separately conserved maps onto a local model of microscopic electrons. We exemplify this through three natural phases for the chargon. 

First, the chargon can form a Mott insulator. A model achieving this is given by the Hamiltonian density
\begin{align}
\label{eq.chargon_Mott}
 H_\text{Mott} &= P_y + P_{y}^\dagger~.
\end{align}
Microscopically, $H_\text{Mott}$ maps onto an on-site density-density interaction for the electrons [cf.~Eq.~\eqref{eq.WA-F.Der.psi-c-terms-1}]. When dominant, this interaction localizes the electron charge. Consequently, the system realizes a pure spin model. In particular, spin-exchange terms are encoded in spinon hopping as described by Eq.~\eqref{spinonhop} with $\vec s_{R/L} \rightarrow \vec N$, the N\'eel vector. The description of spin models in terms of spinons coupled to a compact gauge field was thoroughly analyzed in Ref.~\onlinecite{Leviatan2020}. In particular, various examples of trivial and fractional phases were described in detail.

Alternatively, the chargon can realize a superfluid phase, e.g., when its nearest-wire hopping is dominant. Once $c_y$ spontaneously acquires an expectation value, the gauge field acquires a Higgs mass. Microscopically, nearest-wire hopping for the chargon translates to 
\begin{align}
\label{eq.chargon_cond_micro}
 H_m = \psi^\dagger_{2y+1,\down,L}\psi_{2y,\down,R}+ \psi^\dagger_{2y,\up,R}\psi_{2y-1,\up,L}+\text{H.c.}
\end{align}
The same Hamiltonian was instrumental to the analyses of Mott transitions out of various superconductors in Ref.~\onlinecite{Leviatan2021}. There, it was constructed to permit the countering of inter-wire charge transfer without competing against the superconducting parent Hamiltonian. Here, $H_m$ places half of the long-wavelength electrons into a trivially gapped phase. The remaining electrons are identified with the spinons by the chargon-condensate [cf.~Eq.~\eqref{eq.WA-F.Der.psi-c-f}]. Notice that the local electron model $H[\psi_\sigma]$ obtained from $H_m[c]+H'[f_\sigma]$ does not microscopically coincide with $H'[\psi_\sigma]$. However, we expect $H[\psi_\sigma]$ and $H'[\psi_\sigma]$ to realize the same phase. This property was shown for an $s$-wave superconductor in Ref.~\onlinecite{Leviatan2021} and is demonstrated for the $K=n$ state in Appendix \ref{App.B}.

The chargon can also form fractional phases. For example, a $\nu=\frac{1}{2 n}$ Laughlin state for $c_y$ arises when the interaction
\begin{align}
\label{eq.chargon_QH}
 H_\text{QH} &=c^\dagger_{y+1} (P_{y+1} P_y)^n c_y +\text{H.c.} ~
\end{align}
dominates. The gauge field $a$ becomes massive due to an induced Chern-Simons action. The spinons are then composite fermions. Specifically, they are related to the microscopic electrons by the attachment of $4 \pi n$ flux independent of their spin. They are natural variables for describing various spinful quantum Hall states. Coupled-wire models $H_\text{KML}[\psi_\sigma]$ for such phases have been previously constructed, following a different philosophy, in Refs.~\onlinecite{Kane2002,TeoKane2014}. 

\section{Discussion}\label{discussion}
We have analyzed spin-1/2 particles on a square lattice near half-filling within a vortex formulation. Taking the long-wavelength limit within the dual theory, we found a dynamical decoupling of electrons into charge-neutral spinons and spinless chargons. The latter encode the emergent degrees of freedom at various unconventional phase transitions, such as between superconductors and valence bond solids. In addition, they readily capture various phases with fractional excitations. As our analysis is tailored to nearly half-filled systems, it takes into account the most important lattice effect at the leading order. The impact of additional perturbations, such as phonons, disorder, or spin-orbit coupling, can be readily deduced from within the theory, based on symmetries and relevance under renormalization.

Similar decompositions into spinons and chargons are often asserted on the lattice scale as the starting point of parton approaches. These are more general than the dual formulation and can access a broader range of phases. In particular, parton field theories are not restricted to systems that are close to the Mott transition. They thus provide little guidance for finding microscopic systems that realize any given parton ansatz. By contrast, our dual formulation utilizes only the long-wavelength degrees of freedom near the Mott transition. Consequently, we expect all phases discussed here to be accessible by weakly perturbing systems near such a transition.

The narrower applicability of the dual description is offset by retaining more microscopic information. In coupled-wire systems, the duality even provides an exact mapping between microscopic variables and gauge-theory quantities. More generally, we expect the dual approach to be suitable for properties that do not directly relate to symmetries and thus difficult to incorporate in parton-based field theories. A prominent example of such a property may be geometric frustration. In its absence, on a square lattice, we found that the long-wavelength structure leads to spinons at integer filling. Consequently, it does not readily permit spinon Fermi surface states. Such states have been proposed, in particular, for triangular lattices.\cite{Lee2005,Motrunich2005} It would thus be very interesting to perform a similar analysis there.

Our considerations apply to systems that conserve charge and one spin component. We do not explicitly retain the full $SU(2)$ spin-rotation symmetry, which would be a fruitful avenue for future work. Instead, the charge and the conserved spin component are treated on equal footing. Interchanging them leads to an alternative decomposition of electrons into a single spinon and a pair of chargons. Similarly, it maps certain phases and transitions onto each other.

Another fruitful direction is the exploration of one-dimensional analogues of the systems described here. Ref.~\onlinecite{jiang2019} showed that the transition between an Ising magnet and VBS in one dimension shares many features of the N\'eel-VBS transition with continuous spin-rotation symmetry in two dimensions. Their finding is a non-trivial manifestation of the close correspondence between the one-dimensional Kramers-Wannier and two-dimensional boson-vortex dualities.\cite{Senthil2019,Karch2019} We expect that a similar analogue exists for SCS-duality presented here. Its study may provide an interesting perspective on phase transitions in one dimension and valuable lessons for two-dimensional models.

Finally, we complemented the field-theoretic duality with an exact microscopic implementation within the coupled-wire framework. It readily yields parent Hamiltonians for numerous fractional phases and the transitions between them. Specifically, one can construct concrete models where the charge (spin) response changes qualitatively, but the spin (charge) gap remains. Such wire constructions are readily translated to lattice Hamiltonians that may then be studied numerically.

\begin{acknowledgments}
This work was supported by the Deutsche Forschungsgemeinschaft (CRC/Transregio 183) and the 
Israel Science Foundation (2572/21).
\end{acknowledgments}

\bibliography{chargepartons.bib}

\appendix
\section{Coupled-wire derivation of SCS-duality}
\label{App.A}
Here we derive the SCS-duality in an array of one-dimensional systems, i.e., wires. Each wire, labeled by the index $y$ and extending along $\hat x$, hosts spinful bosons $B_{y,\sigma}$ at an average filling of $\rho^0_{\sigma} = \frac{1}{2}$. We use Abelian bosonization\cite{GiamarchiBook2004} to encode the bosonic long-wavelength fluctuations through pairs of conjugate fields $\Phi_{y,\sigma}$ and $\Theta_{y,\sigma}$ with the convention
\begin{align}
 B_{y,\sigma} = e^{i\Phi_{y,\sigma}}~, \qquad
 \rho_{y,\sigma} = \rho^0_{\sigma}+ \frac{1}{\pi}\partial_x\Theta_{y,\sigma}~.
\end{align} 
Here $\rho_{y,\sigma}$ is the total density operator for spin-$\frac{\sigma}{2}$ bosons. Each step in the field-theoretic derivation of the SCS-duality corresponds to a specific, non-local mapping for the bosonized operators.\cite{Mross2016,Mross2017,Fuji2019,Leviatan2020}

\subsubsection*{Step 1: First duality and low-energy vortices}
The creation operator of a vortex $V_{\sigma} \equiv e^{i\tilde{\Phi}_{\sigma}}$, and the corresponding density $\rho^V_\sigma=\frac{1}{\pi}\partial_x\tilde{\Theta}_{\sigma}$ are expressible through the conjugate variables~\cite{Mross2017}
\begin{subequations}
\label{eq.BosonVotex}
\begin{align}
 \label{eq.BosonVotex_phi}
 \tilde{\Phi}_{\tilde{y},\sigma} &= \sum_{y'} \text{sgn}(y'-\tilde{y})\Theta_{y',\sigma}~,\\
 \label{eq.BosonVotex_theta}
 \tilde{\Theta}_{\tilde{y},\sigma} &= \tfrac{1}{2}(\Phi_{y+1,\sigma}-\Phi_{y,\sigma})~,
\end{align}
\end{subequations}
with $\tilde{y}=y+\frac{1}{2}$ denoting dual wires. In the wire framework, it is most convenient to work in a gauge where vortex hopping along the wire is translation invariant. As a consequence, the two-degenerate valleys in the vortex band structure arise at momenta $k_y=0,\pi$. Even and odd linear combinations of the low-energy vortices then each reside on dual wires with a specific parity. Specifically, we identify
\begin{equation}
\begin{alignedat}{2}
 v_{\up,1}&= V_{2\tilde{y}-1,\up}~, \qquad \qquad && v_{\down,1}= V_{2\tilde{y},\down}~,\\
 v_{\up,2}& = V_{2\tilde{y},\up}~, \qquad \qquad && v_{\down,2}= V_{2\tilde{y}-1,\down}~.
\end{alignedat}
\end{equation}
The operator $M_\sigma$ defined in Sec.~\ref{sec.SCS_duality} translates to
\begin{align}
 M_{y,\sigma}& = e^{i\sigma(-1)^y (\tilde\Phi_{\tilde{y},\sigma} - \tilde\Phi_{\tilde{y}-1,\sigma})} = e^{-i\sigma(-1)^y 2\Theta_{y,\sigma}} \label{eqn.bosoizedmonopoles},
\end{align}
and is odd under translations along $x$. The operator $e^{i2\Theta^{c}}$ with $\Theta^{c}=(\Theta_\uparrow +\Theta_\downarrow)$ introduce a $2\pi$ phase slip into both spin species and is even under translations. Eq.~\eqref{eqn.bosoizedmonopoles} identifies it with the operator $M_s$ of Sec.~\ref{sec.SCS_duality}. Similarly, $e^{i2\Theta^{s}}$ with $\Theta^{s}=(\Theta_\uparrow -\Theta_\downarrow)$ corresponds to the operator $M_c$. Notice that $C_4$ rotations are necessarily broken in the wire setup. The operators $M_{c,s}$ are both even under translations and $C_2$ rotations.

The low energy vortices couple minimally to two emergent, non-compact gauge fields $\tilde{a}_{\tilde{y}}^{\sigma}$ governed by Maxwell dynamics.\cite{Mross2017} We express the four low-energy vortices via
\begin{subequations}
\begin{align}
 \vec{\tilde \Phi}& = \begin{pmatrix}\tilde \Phi_{2,\up} & \tilde \Phi_{2,\down} & \tilde \Phi_{1,\up} & \tilde \Phi_{1,\down}\end{pmatrix}^T~, \\
 \vec{\tilde \Theta}& = \begin{pmatrix}\tilde \Theta_{2,\up} & \tilde \Theta_{2,\down} & \tilde \Theta_{1,\up} & \tilde \Theta_{1,\down}\end{pmatrix}^T~,
\end{align}
\end{subequations}
and find that $\tilde a^\sigma$ couple to the vortex densities via
\begin{align}
 \mathcal{L}_\text{coupling} =\frac{i}{\pi}\sum_{\sigma}\partial_x \vec{\tilde{\Theta}}\cdot \vec{Q}_\sigma \tilde{a}_{0}^{\sigma}~,
\end{align}
with charges $\vec Q_\up=\begin{pmatrix} 1 & 0 & 1 & 0 \end{pmatrix}$ and $\vec Q_\down = \begin{pmatrix} 0 & 1 & 0 & 1 \end{pmatrix}$. 

\subsubsection*{Step 2: Reparametrization and hybridization of vortices}
Following Sec.~\ref{sec.SCS_duality}, which introduced $v_{b_\sigma}=v_{\sigma,2}$ and $v_{d_\sigma}=v_{\sigma,1}v_{-\sigma,2}$, we define their coupled-wire implementations $v_{b_\sigma} \equiv e^{i\tilde\varphi^b_{\sigma}}$ and $v_{d_\sigma} \equiv e^{i\tilde\varphi^d_{\sigma}}$. The phase variables are conveniently expressed through the vector $\vec{\tilde \varphi} = \begin{pmatrix}\tilde \varphi^b_{\up} & \tilde \varphi^b_{\down} & \tilde \varphi^d_{\up} & \tilde \varphi^d_{\down}\end{pmatrix}^T$, which is related to $ \vec{\tilde \Phi}$ as
\begin{equation}
\label{eqnparamter}
 \vec{\tilde \varphi} \equiv W \vec{\tilde \Phi} ; \qquad W = \begin{pmatrix} 
 1 & 0 & 0 & 0 \\
 0 & 1 & 0 & 0 \\
 0 & 1 & 1 & 0 \\
 1 & 0 & 0 & 1
 \end{pmatrix}~.
\end{equation}
The conjugate fields, given by $\vec{\tilde \theta} \equiv W^{-T}\vec{\tilde \Theta}$ encode the corresponding densities. Consequently, $v_{b_\sigma},v_{d_\sigma}$ couple to the gauge fields $\tilde a^\sigma$ with charges $\vec q = W \vec Q$, in agreement with Eq.~\eqref{eq.low_e_vortices-d}. In terms of the new vortices we find $M_s = v_{d_\up}^\dagger v_{d_\down}$. In its presence, $v_{d_\sigma}$ hybridize. We, therefore, define a single vortex $v_c=e^{i\tilde \varphi^c}$ with 
\begin{align}
 \tilde \varphi^c_{\tilde{y}} = \begin{cases}
 \tilde \varphi^d_{\tilde{y},\up}~, & \tilde{y} \text{ odd}~, \\
 \tilde \varphi^d_{\tilde{y},\down}~, & \tilde{y} \text{ even}~,
 \end{cases}
\end{align}
and the corresponding density $\tilde\rho^c = \frac{1}{\pi}\partial_x\tilde \theta^c$.

\subsubsection*{Step 3: Second duality yielding spinons and chargons}
We complete the derivation by introducing the chargon $c=e^{i\varphi^c}$ as the vortex in $v_c$ and the spinons $b_\sigma = e^{i\varphi^b_\sigma}$ as vortices in $v_{b_\sigma}$. Their density fluctuations are given by $\delta\rho^c=\frac{1}{\pi}\partial_x\theta^c$ and $\delta\rho^b_\sigma=\frac{1}{\pi}\partial_x\theta^b_\sigma$, respectively. The new fields follow relations similar to Eq.~\eqref{eq.BosonVotex}, i.e.,
\begin{subequations}
\begin{align}
 \varphi^c_{y} &= \sum_{y'} \text{sgn}(y-\tilde{y}')\tilde\theta^c_{\tilde{y}'}~, \\
 \theta^c_{y} &= \tfrac{1}{2}(\tilde\varphi^c_{\tilde{y}-1}-\tilde\varphi^c_{\tilde{y}})~,\\
 \varphi^b_{2\tilde{y}+1,\up} &= -\sum_{\tilde{y}'} \text{sgn}(2\tilde{y}-2\tilde{y}'-1)\tilde\theta^b_{2\tilde{y}',\up}~, \\
 \theta^b_{2\tilde{y}+1,\up} &= \tfrac{1}{2}(\tilde\varphi^b_{2\tilde{y},\up}-\tilde\varphi_{2\tilde{y}+2,\up}^b)~,\\
 \varphi^b_{2\tilde{y},\down} &= -\sum_{\tilde{y}'} \text{sgn}(2\tilde{y}-2\tilde{y}'-1)\tilde\theta^b_{2\tilde{y}'-1,\down}~, \\
 \theta^b_{2\tilde{y},\down} &= \tfrac{1}{2}(\tilde\varphi^b_{2\tilde{y}-1,\down}-\tilde\varphi_{2\tilde{y}+1,\down}^b)~.
\end{align}
\end{subequations}
These relations, combined with Eq.~\eqref{eqnparamter} and Eq.~\eqref{eq.BosonVotex} yield non-local expressions for individual chargon and spinon creation operators in terms of the microscopic bosons. By contrast, the densities and hopping operators of spinon and chargons are all local microscopic operators. They are given by
\begin{subequations}
\begin{align}
 &\varphi^c_{y+1}-\varphi^c_{y} = \begin{cases}
 \Phi_{y+1,\up} - \Phi_{y,\up}~, & y \text{ odd}~, \\
 \Phi_{y+1,\down} - \Phi_{y,\down}~, & y \text{ even}~,
 \end{cases} \\
 &\theta^c_y = \Theta_{y,\up} + \Theta_{y,\down}~,
 \\
 &\varphi^b_{\tilde{y}+1,\sigma}-\varphi^b_{\tilde{y},\sigma} = \Phi_{y+1,\sigma} - \Phi_{y+1,-\sigma}- \Phi_{y,\sigma} + \Phi_{y,-\sigma}~,
 \\
 &\theta^b_{\tilde{y},\sigma} = \Theta_{y+1,\sigma}+\Theta_{y,\sigma}~.
\end{align}
\end{subequations}
The creation operators of microscopic bosons are represented as
\begin{subequations}
\begin{align}
 B_{2y,\sigma} &= c_{2y} b_{2y-\tfrac{\sigma}{2},\sigma}~,\\
 B_{2y+1,\sigma} &= c_{2y+1} b_{2y+1+\tfrac{\sigma}{2},\sigma}~.
\end{align}
\end{subequations}

\subsubsection*{Step 4: Fermions}
\label{app.A.f}
We turn to a microscopic wire array that hosts spinful fermions at an average filling of $\rho^0_\sigma = \frac{1}{2}$. We express the long-wavelength fermions within the framework of Abelian bosonization as $\psi_{y,\sigma,\chi}=e^{i(\Phi_{\psi,\sigma}+\chi\Theta_{\psi,\sigma})}$, with $\chi=R/L$. Their density fluctuations are encoded as $\delta\rho_{ y ,\sigma,}\equiv \frac{1 }{\pi}\partial_x \Theta_{ y ,\sigma} $. The flux attachment of Eq.~\eqref{eq.LCS} corresponds to the operator mapping 
\begin{subequations}
\label{eq.wire_FA}
\begin{align}
 \Phi_{\psi,y,\sigma} &= \Phi_{y,\sigma} + \sigma \sum_{y'}\sgn{y'-y}\Theta_{y',\sigma}~,\label{eq.wire_FAa}\\
 \Theta_{\psi,y,\sigma} &= \Theta_{y,\sigma}~.
\end{align}
\end{subequations}
We use the same flux attachment to express the fermionic spinons in terms of the bosonic ones (cf.~Sec.~\ref{sec.fermionscs}). The operator mapping is the same as in Eq.~\eqref{eq.wire_FA} with
\begin{equation}
\begin{alignedat}{2}
 &\Phi_{\psi,y,\sigma}&&\rightarrow\varphi_{f,\tilde{y},\sigma}~,\\
 &\Phi_{y,\sigma}&&\rightarrow\varphi_{\tilde{y},\sigma}~,\\
 &\Theta_{\psi,y,\sigma}&&\rightarrow\theta_{f,\tilde{y},\sigma}~,\\
 &\Theta_{y,\sigma}&&\rightarrow\theta_{\tilde{y},\sigma}~,
\end{alignedat}
\end{equation}
and where for $\sigma=\text{$\up$($\down$)}$ the sum in Eq.~\eqref{eq.wire_FAa} only includes odd(even) $\tilde{y}'$. Finally, chiral spinons are expressed as $f_{\tilde{y},\sigma,\chi} \equiv e^{i(\varphi_{f,\tilde{y},\sigma} +\chi \theta_{f,\tilde{y},\sigma})}$. The operator mappings of the bosonic SCS, along with the flux attachment identities, relate microscopic fermionic operators and those describing the emergent chargons and spinons. These mappings are listed in Sec.~\ref{sec.wirerealization}.

\section{Stitching together two different coupled-wire models}
\label{App.B}
We demonstrate here that two distinct coupled-wire models for the $K=n$ state of fermions realize the same phase. Specifically, we consider the coupled-wire model $H_{\text{KLM}}$ as introduced in Refs.~\onlinecite{Kane2002,TeoKane2014} and $H_{\text{SCS}}$ obtained from translating $H_m[c]+H_{\text{KML}}[f_\sigma]$ to microscopic variables. We follow the bosonization conventions adopted in step 4 of Appendix \ref{App.A}. Dropping the index $\psi$ to lighten the notation, we define $\phi_{\chi,\sigma}\equiv\Phi_{\psi,\sigma}+\chi\Theta_{\psi,\sigma}$.

The first model is conveniently expressed through the variables\cite{TeoKane2014} 
\begin{subequations}
\begin{align}
 \widetilde{\phi}_{y,R,\sigma} &\equiv \frac{n+1}{2}\phi_{y,R,\sigma} -\frac{n-1}{2}\phi_{y,L,\sigma}~,\\
 \widetilde{\phi}_{y,L,\sigma} &\equiv \frac{n+1}{2}\phi_{y,L,\sigma} -\frac{n-1}{2}\phi_{y,R,\sigma}~.
\end{align}
\end{subequations}
It is then given by
\begin{align}
 H_{\text{KML}} =& \sum_\sigma\cos[\widetilde{\phi}_{y+1,L,\sigma}-\widetilde{\phi}_{y,R,\sigma}].
\end{align}
A semi-infinite array of wires with $y>0$, say, exhibits fractional edge modes described by $\widetilde{\phi}_{1,L,\sigma}$. For $H_\text{SCS}$ we analogously introduce
\begin{subequations}
\begin{align}
 \overline{\phi}_{2y,L,\sigma} &\equiv \frac{n+1}{2}\phi_{2y,L,\sigma} -\frac{n-1}{2}\phi_{2y-\sigma,R,\sigma}~,\\
 \overline{\phi}_{2y,R,\sigma} &\equiv \frac{n+1}{2}\phi_{2y-\sigma,R,\sigma} -\frac{n-1}{2}\phi_{2y,L,\sigma}~,
\end{align}
\end{subequations}
and find
\begin{align}
 \nonumber H_{\text{SCS}} =& \sum_\sigma \cos[\phi_{2y,R,\sigma}-\phi_{2y-\sigma,L,\sigma}] \\
 &+ \sum_\sigma \cos[\overline{\phi}_{2y+2,L,\sigma}-\overline{\phi}_{2y,R,\sigma}]~.
\end{align}
A semi-infinite array of wires with $y\leq 0$ exhibits fractional edge modes described by $\overline{\phi}_{0,R,\up},\overline{\phi}_{-2,R,\down}$. In this termination, the $\down$ species additionally exhibits an unprotected pair of counter-propagating modes, which we express by $\widetilde{\phi}_{0,\chi,\down}$.

The two semi-infinite wire arrays can be stitched together by the local interaction
\begin{align}
 \nonumber \delta H = & \cos[\widetilde{\phi}_{1,L,\up} - \overline{\phi}_{0,R\up}] \\
 \nonumber & + \cos[\widetilde{\phi}_{1,L,\down}-\widetilde{\phi}_{0,R,\down}] + \cos[\widetilde{\phi}_{0,L,\down} - \overline{\phi}_{-2,R\down}]\\
 \propto & (\psi_{1,L,\up}\psi_{-1,R,\up}^{\dagger})^{\tfrac{n+1}{2}} (\psi_{1,R,\up}^{\dagger} \psi_{0,L,\up})^{\tfrac{n-1}{2}}\\
 \nonumber & + (\psi_{1,L,\down}\psi^\dagger_{0,R,\down})^{\frac{n+1}{2}}(\psi_{1,R,\down}^\dagger \psi_{0,L,\down})^{\frac{n-1}{2}}\\
 \nonumber & + (\psi_{0,L,\down}\psi_{-1,R,\down}^\dagger)^{\frac{n+1}{2}}(\psi_{0,R,\down}^\dagger\psi_{-2,L,\down})^{\frac{n-1}{2}} + \text{H.c.}
\end{align}
For odd $n$, all exponents are integers and the interaction $\delta H$ is microscopically allowed. It results in a fully gapped array and permits fractional excitations to traverse the boundary between $y<0$ and $y>0$. Consequently, $H_\text{KML}$ and $H_\text{SCS}$ realize the same topological phase.

\end{document}